# A theory of gravity as a pressure force

# II. Lorentz contraction and "relativistic" effects

M. Arminjon, Grenoble, France

**Summary.** In a foregoing paper, gravity has been interpreted as the pressure force exerted on matter at the scale of elementary particles by a perfect fluid. Under the condition that Newtonian gravity must be recovered in the incompressible case, a scalar field equation has thus been proposed for gravity, giving a new theory in the compressible case. Here the theory is reinterpreted so as to describe the relativistic effects, by extending the Lorentz-Poincaré interpretation of special relativity which is first recalled. Gravitational space-contraction and time-dilatation are postulated, as a consequence of the principle of local equivalence between the effects of motion and gravitation. The space-time metric (expressing the proper time along a trajectory) is hence curved also in the proposed theory. As the result of a modified Newton law, it is proved that free test particles follow geodesic lines of this metric. In the spherical static situation, Schwarzschild's exterior metric is exactly recovered and with it the experimental support of general relativity, but the interior solution as well as the problematic of singularities are different in the proposed theory, e.g. the radius of the body cannot be smaller than the Schwarzschild radius.

1. INTRODUCTION

In a foregoing paper [1] a new, scalar theory of gravitation has been proposed within purely classical concepts of space and time. In order to obtain a more complete theory, it is necessary to have a description of the "relativistic" effects, i.e. the effects of motion and gravitation on the measurement of distances and time intervals, the dependence of inertial mass on the velocity, and the mass-energy equivalence. These effects are so called because they are unified within, and several of them were first predicted by Einstein's special and general theories of relativity, but the effect of uniform motion on the distances and times, as well as the mass increase with the velocity, were first predicted as "absolute" effects in Lorentz's theory of ether ; the appearance of a "local" time in a uniformly moving frame was suggested by Poincaré and defined by him as empirically relevant [2]. The ether postulated here, seen at a macroscopic scale, is suitable for the modern version of the Lorentz-Poincaré electromagnetic ether, developed by Builder [3], Prokhovnik [4] and others. This means that all the results of Einstein's *special* relativity will be available in the corresponding "absolute" interpretation which will be briefly recalled. Of course, in the presence of a gravitational field, special relativity holds only locally, i.e. here in domains where the (macro-) ether pressure may be considered as uniform. Then an analysis of the "equivalence principle", which concerns here the equivalence between the local effects of a motion with respect to the ether and those of a gravitational field i.e. a field of ether pressure, will lead to postulate a gravitational space-contraction and time-dilatation. As a consequence, the equations of the proposed gravitation theory are reinterpreted in terms of the physical space metric (a riemannian one) instead of the euclidean metric. After an analysis of the status and meaning of a space-time metric and of Einstein's geodesic assumption in the proposed theory, the static problem with spherical symmetry is reexamined. Schwarzschild's exterior space-time metric (and thus the experimental support of general relativity) is exactly recovered in the proposed theory, but this is not the case for the interior metric, and also the question of singularities is not posed in the same terms.

## 2. THE LORENTZ-POINCARE INTERPRETATION OF SPECIAL RELATIVITY

The decisive change from classical to modern physics came when it was realized that the need for operative definitions of physical quantities also applies to space and time. Since this idea has been expressed with a great lucidity by Einstein and since his theory gives the first place to the requirement of frame indifference, this essential change is generally known as Einstein's relativity. It is important for our purpose, however, to recall that the precursors of Einstein (who were not less than Lorentz and Poincaré) derived a significant part of the results of special relativity *within a theory based on ether*. Even the synchronization of the clocks in a rigid frame in uniform motion (relative to the ether) with the help of light signals was first introduced by Poincaré in 1900, as corresponding, up to the first order in v, to a "local time" (i.e. in the moving frame) [2] :

$$t' = t - \frac{v\,x}{c^2} \quad . \quad (1)$$

Here t is the "absolute time" (i.e. in the rigid ether), v is the constant velocity of the moving frame, the axis of the abscissa x being in the direction **v**, and c is the light velocity. Up to the first order in v, Eq. (1), which is given in [2] without proof, is the well-known time-component of Lorentz's transformation. This latter was yet introduced by Lorentz in 1904 only, and independently by Einstein in 1905. The Lorentz-Poincaré ether theory, which at their time was less convincing than Einstein's special relativity, has been rigorously presented and developed by Builder [3] and Prokhovnik [4] with the same logical thrift that has made the success of Einstein's theory. The two main assumptions of the latter are well-known : (a) the physical equivalence of all inertial frames, and (b) the constancy of c, in particular its invariance by change of the inertial frame (the first assumption contains in fact the program of changing usual kinematics and newtonian mechanics so that at the same time they keep, as before, the same form in all inertial frames, and Maxwell's electrodynamics also wins this invariance property).

The assumptions of Prokhovnik [4] are : (i) the existence of a reference frame E in which the energy-propagation is isotropic and Newton's second law holds in the form

$$\mathbf{F} = \frac{d\mathbf{P}}{dt} \quad , \quad \mathbf{P} = m\,\mathbf{v} \quad , \quad (2)$$

without assuming, however, that the inertial mass m does not depend on v (note that the relativistic assumption (a) is stronger, since it actually demands that (2) holds in any inertial frame) ; and (ii) the Fitzgerald-Lorentz contraction, understood as a *true* contraction of all material objects after they have been transported from E to a frame $E_\mathbf{u}$ having uniform and constant velocity **u** with respect to E, in a ratio β assumed to depend only on u = |**u**|. This occurs only on lines parallel to **u**, in the sense that a rod AB at rest in $E_\mathbf{u}$ and making the angle θ with **u**, has the length *l'* with

$$(l' \cos \theta / \beta)^2 + (l' \sin \theta)^2 = l^2 \quad , \qquad (3)$$

when θ and *l'* are measured by an observer in E (who measures the euclidean metric of the physical space, in the sense of [1], §2), and *l* is the length of an identical rod, at rest in E [4]. Thus :

$$l' = \frac{l \beta}{\sqrt{1 - (1-\beta^2) \sin^2 \theta}} \quad . \qquad (4)$$

Consider AB (with a mirror at A perpendicular to AB and the like at B) as a "light clock", its time unit being the interval during which the light goes from A to B back and forth [3-4]. The usual addition formula for the velocities holds, provided that all velocities are defined from the same frame (it is merely the additivity of derivation). If evaluated from E where light propagates at the same speed c in all directions, the velocity of the light ray with respect to the rod AB at rest in $E_\mathbf{u}$ is hence, depending on its direction **AB** or **BA** :

$$c_1 = \sqrt{c^2 - u^2 \sin^2 \theta} - u \cos \theta \, , \quad c_2 = \sqrt{c^2 - u^2 \sin^2 \theta} + u \cos \theta \, . \qquad (5)$$

Thus, as measured with the clock of E, the time period of the light clock at rest in $E_\mathbf{u}$ is:

$$(\Delta t)_\mathbf{u} = l' \left( \frac{1}{c_1} + \frac{1}{c_2} \right) = \frac{2 l \beta}{c (1 - u^2/c^2)} \frac{\sqrt{1 - u^2 \sin^2 \theta / c^2}}{\sqrt{1 - (1 - \beta^2) \sin^2 \theta}} \quad . \qquad (6)$$

Now the negative result of the Michelson-Morley experiment means precisely that this period does not depend on angle θ and thus is a constant in the moving frame $E_\mathbf{u}$. This is true if, and only if one has

$$\beta = \sqrt{1 - \xi} \, , \quad \xi = \mathbf{u}^2 / c^2 \, , \qquad (7)$$

as follows immediately from (6) ( in [4], the value (7) is assumed from the beginning and the "only if" part is not obtained). *The negative Michelson experiment means that light clocks are correct clocks, and this implies the Lorentz contraction.* Moreover, *the time unit of $E_\mathbf{u}$ is "dilated"* : $(\Delta t)_\mathbf{u} = \Delta t / \beta$. Prokhovnik [4] shows precisely, and with a lot of crossed calculations, that these results imply all the kinematics of special relativity in the usual (einsteinian) form, i.e. imply the standard Lorentz transformation and the constancy of c (as this is measured, necessarily on to-and-fro paths, with clocks and rods of any frame in uniform motion $E_\mathbf{u}$). The great improvement with respect to the usual version is the perfectly clear interpretation of the space contraction and time dilatation as absolute effects : the rods of $E_\mathbf{u}$ are *really* shorter, the clocks there go really slower as compared with E. This agrees with modern time-measurements (e.g. that of Hafele & Keating [5] who adopt a frankly "absolute" way of relating their results, although they defend conventional relativity). The relativity of simultaneity is interpreted as a clear consequence of the anisotropic "true" propagation of light, Eq. (5), in a frame that moves with respect to the propagating medium E [3-4]. It concerns only the operative definition of this notion in different frames with the help of light signals and does not forbid to think the absolute simultaneity, which is defined in the time of E.

Until now, Eq. (2) has not been used. Actually the obtainment of relativistic mechanics requires to suppose further that : (iii) if a system of material particles conserves in E its total momentum $\mathbf{P} = \sum \mathbf{P}_i$ and the individual rest masses $m_i(v=0)$, then $\mathbf{P_u} = \sum \mathbf{P_{ui}} = \sum m_i(v_{ui}) \mathbf{v_{ui}}$ is conserved in any frame $E_u$, where $\mathbf{v_u}$ is the velocity defined with rods and clocks of $E_u$ (i.e. by using the relativistic velocity transformation). This assumption expresses the natural requirement that the principle of inertia (in a weaker form than in [1], sect. (3.3)) still holds in presence of the Lorentz contraction : obviously, the relativistic transformation of velocities precludes that the momentum conservation passes from E to $E_u$ if the mass is velocity-independent. Then the well-known relations for the mass and the kinetic energy are obtained in the classical way [4] :

$$m(v) = m(v=0) / \beta(v) \quad , \quad T = \int_{v=0}^{v} \mathbf{F} \cdot \mathbf{dx} = (m(v)-m(0)) c^2 \quad . \quad (8)$$

It is also well-known that this expression of m(v) makes Eq. (2) covariant with respect to Lorentz transformations; hence the approach based on ether gives this result consistently, i.e. the lorentzian equivalence of the frames $E_u$ follows from rather "absolute" assumptions (in [1], sect.(3.3), the same has been found for the galileian equivalence). Thus these frames $E_u$ may be called inertial frames since a test particle has uniform and constant velocity in one of these, $E_{u_0}$ (and hence in all of them) if and only if $\mathbf{F}_{u_0} = 0$. Moreover, the momentum conservation may be postulated for isolated systems also in the case where the individual rest masses are *not* conserved (i.e. in presence of creation or annihilation of particles) : just like in conventional relativity, one finds that the momentum conservation applies in any inertial frame if and only if the total mass of the isolated system is conserved :

$$\sum_{i=1}^{N} m_i(v_i) = \sum_{j=1}^{N'} m'_j(v'_j) \quad . \quad (9)$$

Together with Eq. (8)$_2$, this gives the classical argument of Einstein's special relativity for attributing to any particle the "rest energy" $m(0)c^2$ and identifying mass and energy, up to the factor $c^2$. In a theory based on ether, however, the actio-reactio principle can be postulated in the fundamental frame E (though it does not pass to the moving frames $E_u$ since it is not Lorentz-covariant). This principle is related to the isotropy of physical interactions and is thus consistent with assumption (i). It implies that any isolated system conserves its total momentum, whereas this property is practically a definition of isolated systems in the usual version of special relativity. However a set of material particles is here, strictly speaking, never an isolated system since it is embedded in the ether which is disturbed by the particles [2]. This gives an intuitive version of the classical warning in relativistic textbooks, that the energy of the field has also to be considered. A first disturbance is the dragging of an amount of the fluid ether together with the particle - an amount which should increase with the velocity of the particle [6].

With the here-considered *constitutive* ether, the mass increase with velocity should be a real increase of the amount of ether which defines the material particle. If the particle is a vortex i.e. a special local flow in ether, the "amount" to consider is the kinetic energy in the local flow, divided by $c^2$, thus explaining the mass increase [7]. Then the mass-energy conser-vation for an isolated system of particles (9) would express simply the conservation of the total amount of ether (or kinetic energy, up to the $c^2$ factor) in this set of vortices.

3. EQUIVALENCE PRINCIPLE WITH LORENTZ CONTRACTION

Mechanics and the gravitation theory based on ether have now to be modified so as to take into account the foregoing relativistic effects, which all follow from the Lorentz contraction. The theory [1] introduces the ether as a perfect barotropic fluid, i.e. its pressure depends only on its density, and the frame defined by the mean motion of the ether defines the fundamental inertial frame, which is thus a deformable frame. The gravity acceleration is the transcription of the macroscopic pressure $p_e$ into the force per unit "mass" $\rho_e$ of the macro-ether (i.e. $\rho_e$ is the volume density of the macro-ether):

$$\mathbf{g} = - \frac{1}{\rho_e(p_e)} \text{ grad } p_e \; , \quad (10)$$

and the pressure $p_e$ is assumed to obey the field equation:

$$\Delta p_e - \frac{1}{c_e^2} \frac{\partial^2 p_e}{\partial t^2} = 4\pi G \rho \rho_e \; , \quad p_e = p_e(\rho_e) \; , \quad c_e = \sqrt{\frac{dp_e}{d\rho_e}} \; , \quad (11)$$

where G is Newton's gravitation constant and $\rho$ is the volume density of matter.

As in general relativity, the guide for the modification will be the principle of *local* equivalence between the effects of a gravitational field and those of an "inertial field". Here "local" means : in a domain where the field may be considered as uniform, it corresponds thus to the newtonian equivalence between the effects of inertial and gravitational forces on a mass point (cf. [1], sect. (3.1)). The inertial effects come from the motion with respect to the inertial frame, i.e. the macro-ether E. The important difference with newtonian theory as well as with usual relativity is that the true Lorentz contraction destroys the complete equivalence between E and the other "inertial frames" $E_\mathbf{u}$ with $\mathbf{u} \neq 0$. This means that uniform motion also has "inertial" effects, and since the equivalence between gravitational and inertial fields is local i.e. for uniform fields, these effects are precisely the essential ones. In a theory based on ether we may transpose the effects of uniform motion as such, into effects of a gravitational field (in contrast, general relativity is forced to adopt an indirect way - the famous rotating disc - because in its conventional form special relativity is exactly built to obtain a perfect equivalence between all inertial frames). Now the effects of uniform motion are on the "space-time metric", i.e. concern the behaviour of rods and clocks : more precisely, the rods are contracted in a ratio β in the direction of motion and the clock periods are dilated in the *same* ratio. The principle of equivalence would thus be simultaneously taken into account and physically explained if we would admit that : (iv) *in a gravitational field, the rods are contracted only in the direction of the field* (that of the ether pressure gradient) *and the clock periods are dilated in the same ratio β*. From now on, we investigate the consequences of this assumption, which concerns of course clocks and rods *at rest in E* .

What should be the value of β ? It must obviously depend on the field of macro-ether pressure $p_e$ or equivalently on the field $\rho_e$ , but one has to ask if the relevant field is that which refers (e.g. for the volume evaluation) to the euclidean metric, not affected by motion and gravitation, or that which would be measured with physical rods - the corresponding metric will be called hereafter the *physical metric*. Moreover one has to determine if the field intervenes directly by its local value or by some derivatives. To answer these questions, let us investigate the equivalence more deeply. Consider the ideal situation of special relativity : there is no gravity i.e. no pressure gradient. For an observer at rest in a uniformly moving frame $E_\mathbf{u}$, the macro-ether (which by definition is at rest in E) has a lower density $\rho_{e\mathbf{u}} = \rho_e \cdot \beta(\xi)$ (see Eq. (7)) ; indeed, its "mass" is unchanged (since the mass increase with velocity concerns *material particles*) while its apparent volume increases by the factor $1/\beta(\xi)$, due to the Lorentz contraction of the measuring rod in direction $\mathbf{u}$. Hence the equivalence is perfectly understood, only if β depends merely on $\rho_e$ as evaluated with the physical metric, namely

$$\beta = \rho_e / \rho_e^\infty , \qquad (12)$$

where $\rho_e^\infty$ is the macro-ether density in a region where no gravity is present. This occurs at large distances from massive bodies. Since $\rho_e$ decreases towards the gravitational attraction, we must thus take for $\rho_e^\infty$ the largest value of $\rho_e$ in the (model of) universe; and we must assume that this value is attained (asymptotically) in volume domains, i.e. corresponds to regions where the ether density is indeed uniform, so that the physical metric is indeed euclidean there. In [1], §2, it was assumed that the euclidean metric of the reference manifold coincides with the physical one in "undisturbed regions" ; now we may precise that these correspond to those domains where the physical ether density $\rho_e$ is equal to $\rho_e^\infty$. The transitive ratio β determines the *spatial* variation of the euclidean size of a given measuring rod and of the rate of a given clock, assuming that such equivalent standards exist at any point *bound with ether* (if one forgets this specification, assumption (iv) is meaningless due to the Lorentz space-contraction and time-dilatation). It thus becomes clear that *when accounting for Lorentz and gravitational space-contraction and time-dilatation, the reference body* M *must be the macro-ether itself*, hence eliminating the arbitrary motion of the ether in M. In view of the transitivity of β, always the same variation of the rate of clocks will be predicted by Eq. (12) (with $\rho_e^0 = \rho_e(\mathbf{x}_0)$ instead of $\rho_e^\infty$) if one selects any arbitrary point $\mathbf{x}_0$ of ether. But in order to deduce the relation between the physical space metric $\mathbf{g}$ and the euclidean metric $\mathbf{g}^0$ from assumption (iv) we must take the point $\mathbf{x}_0$ outside the gravitation field $\mathbf{g}$, i.e. $\rho_e^0=\rho_e^\infty$, so that $\mathbf{g}$ coincides with $\mathbf{g}^0$ at $\mathbf{x}_0$ : otherwise it would be impossible to determine two physical directions orthogonal to $\mathbf{g}(\mathbf{x}_0)$ according to $\mathbf{g}^0$ and to apply assumption (iv). In treatises on general relativity, the gravitational time-dilatation is evaluated early in the construction of the theory (well before obtaining Einstein's equations), at least for weak fields [8],[11]. This is based only on Einstein's assumption that test particles follow the geodesic lines of the *physical* space-time metric. Recently, it has been proved by Mazilu [9] that already in pure newtonian theory one can *build* a space-time metric, the geodesic lines of which are exactly the newtonian trajectories. In the quoted treatises, it is shown that if U is the potential of a weak newtonian field (**g** = grad U with U<<$c^2$, the arbitrary constant being so determined that U cancels at large distances from massive bodies), the geodesic assumption gives the slowing up :

$$\beta = \frac{d\tau}{dt} = \sqrt{1 - \frac{2U}{c^2} + O(\frac{1}{c^3})} \approx 1 - \frac{U}{c^2} < 1 \quad , \quad (13)$$

where t is the time measured outside the field (far from the masses) and $\tau$ is that measured with the dilated clock period in the field. In our theory, a weak gravitational field derives from the potential ([1], Eq. (38b)) :

$$U = -c_e^2 \text{Log} \frac{\rho_e}{\rho_e^\infty} \approx c_e^2 \frac{\rho_e^\infty - \rho_e}{\rho_e^\infty} \quad ,$$

$$1 - \frac{U}{c_e^2} \approx \frac{\rho_e}{\rho_e^\infty} \quad . \quad (14)$$

Until now, nothing has been assumed for the value of $c_e$, the "sound" velocity in the compressible ether. Clearly, if material particles are ether vortices, one may expect that their velocity is limited precisely by $c_e$ (since they should be destroyed by shock waves if exceeding this velocity). On the other hand, the relativistic mechanics implied by the Lorentz contraction give a different limit, namely the light velocity c (Eqs.(7) and (8)). Thus *we assume $c_e = c$*, and Eqs. (12)-(14) show that assumption (iv) gives the same slowing up of the clocks than does the geodesic assumption (for weak fields).

4.REINTERPRETATION OF THE FIELD EQUATION FOR ETHER PRESSURE

When applied now to the space contraction, the value (12) for $\beta$ turns out to imply that *as evaluated with respect to the euclidean metric, the ether density is uniform* (indeed, an amount $dm_e$ of ether occupies the volume $dV = dm_e/\rho_e$ when this volume is evaluated with physical rods, contracted in the direction of gravitational attraction : thus this volume is $dV_0 = dV \beta = dV \rho_e / \rho_e^\infty = dm_e / \rho_e^\infty$ as evaluated with the euclidean metric). Since our gravitation theory assumes a finite ether compressibility, this means that *when accounting for the Lorentz contraction, the equations for the field of ether pressure* (10)-(11) (and [1], sects. (4.1) and (4.2)) *must be written in terms of the physical metric*. Is that possible? First, it is easy to see that all these equations have a sense, since they involve the div, grad and rot operators (along with combinations) which may be defined for a 3-D riemannian metric as well [10]. However the time which appears in some of them has to be defined, because with the slowing down of clocks in the field, one may either use the local time (of clocks at rest in the ether, but in the field) or the "absolute" time which may be defined as that of a clock at rest in the ether *and* far from massive bodies (i.e. in those regions where $\rho_e=\rho_e^\infty$). Since these equations are local and are intended to relate physical effects, observable locally, we assume that *the time also is local in the equations* (in the case of a motion in the ether, one thus has to use the time $t_x$ of a clock that momentarily coincides with the moving point **x**(s) ; the chain rule implies that the time derivatives of order n involve all derivatives with respect to the parameter s on the trajectory, up to the order n). Now in small domains where $\rho_e$ may be considered as uniform, the physical space metric, as evaluated by an observer bound with the ether, is an euclidean one, since it is deduced from the euclidean metric of the reference manifold M by a point-dependent orthogonal affinity (anyhow, a riemannian metric becomes euclidean in the infinitesimal - but here one gets a feeling of what is infinitesimal : the spatial variation of $p_e$ must be small as compared with $p_e$). Thus we can take it for granted that *the mechanics of special relativity hold true in these small domains*, in the interpretation based on ether (§ 2) ; in particular Newton's second law applies with Einstein's modification (Eqs. (2) and (8)$_1$). In static situations this amounts to usual newtonian theory : hence all the arguments of [1], sects. (4.1) and (4.2), apply equally with the riemannian physical space metric, up to (but not including) the discussion of the unsteady situation ; one just has to note that the requirement of regaining newtonian theory for an incompressible ether remains as such, because with an incompressible ether the euclidean metric is regained at the same time than Poisson's equation. One may now raise the question of how far the gravitation field **g** (Eq. (10)) can still be considered as a force per unit (passive gravitational) mass, acting on *moving* test particles in the sense of Eq. (2), and if the rest mass or the velocity-dependent mass must be multiplied by **g**. Clearly, *if Newton's second law may still apply, the time variation of the momentum must be evaluated with the time $t_x$ of the momentarily coinciding clock defined above, and with respect to the non-euclidean physical space metric (see Appendix)*. As far as the velocity **u**=d**x**/dt$_x$ of the test particle is small enough, one can neglect the contribution of the variation in $\rho_e$ along the trajectory (d$\rho_e$/dt$_x$ = **u**.grad $\rho_e$), i.e. the variation in both the physical metric and the rate of the coinciding clock ; then Eq. (2) applies in the usual sense, also with the gravitational force - though the permissible velocity range is reduced when the latter increases. It is clear also that the gravitational force must be **F** = m(u) **g**, with u=|**u**| and m(u) from Eq. (8), in order to save the identity between inertial and passive gravitational mass. This is consistent with the interpretation of gravity as the macroscopic pressure action in the constitutive ether, since this interpretation demands that the mass density $\rho_p$ of the elementary particles (or at least their average density) coincides with the macro-ether density $\rho_e$ ([1], Eqs. (31)-(32)). Since the mass increase with velocity is interpreted as a real increase of the amount of ether constituting the material particles (or of the kinetic energy of the corresponding vortices)(§1), the mass density $\rho_p$ to be considered must be the velocity-dependent one : then the pressure force acting on a (set of) moving material particle(s) is indeed (Eq. (10)) :

$$\mathbf{F}(u) = - V(u) \, \text{grad} \, p_e = V(u) \, \rho_e \, \mathbf{g} = V(u) \, \rho_p(u) \, \mathbf{g} = m(u) \, \mathbf{g} \quad . \quad (15)$$

Assuming that $\rho_e = \rho_p$ is consistent with the very low ether compressibility,
K = d$\rho_e$ / dp$_e$ = 1/ c$^2$. Thus our modified Newton law writes :

$$\mathbf{F} \equiv \mathbf{F}_0 + m(u) \, \mathbf{g} = \frac{d}{dt_x}\left( m(u) \frac{d\mathbf{x}}{dt_x} \right) , \quad \frac{d}{dt_x} \equiv \frac{\rho_{e0}}{\rho_e} \frac{d}{dt_{x_0}} \quad , \quad (16)$$

where $\mathbf{F}_0$ is the non-gravitational force and $x_0$ is a fixed point of ether with density $\rho_{e0}$.

Furthermore, it is natural to admit that, as in newtonian theory, the active gravitational mass equals the passive one. The former is the source of the disturbance in the ether pressure or more formally the mass, the density $\rho$ of which enters the right-hand side of Eq. (11). This assumption may be considered, also in our theory, as a consequence of the actio-reactio principle which has been assumed to remain valid in the rest frame E of the macro-ether (§2). Indeed, whether active or passive, the gravitational force is the transcription of the pressure force into a volume force through the assumption $\rho_e = \rho_p$. The crucial consequence is that *$\rho$ in the equations for the field $p_e$ is the mass-energy density* (relative to the frame E), in other words : *light and* (kinetic) *pressure also produce gravitation*. This is a well-known novelty of general relativity ; in our theory this result follows simply from the interpretation of material particles, including photons, as special local flows (vortices) in the fluid ether, the mass or energy (up to the $c^2$ factor) being the amount of ether or kinetic energy in the considered flow.

The discussion of the unsteady situation can hardly be transcribed in the same way. As has been stated in [1], §2, the continuity equation may be derived also for a riemannian manifold, and even if its metric varies. More precisely, let M be an orientable differentiable manifold, $\mathbf{g}_t$ a metric on M, depending smoothly on the parameter or time t, $\psi_t$ a time-dependent diffeomorphism of M and $\rho_t$ a time-dependent scalar function defined on M, which is "conserved in material volumes" i.e. for any volume domain $\Omega$ :

$$\frac{d}{dt} \int_{\psi_t(\Omega)} \rho_t \, dV_t = 0 \qquad (17)$$

where $V_t$ is the volume defined with the metric $\mathbf{g}_t$. Then the velocity field, defined in the eulerian description as the chart-independent vector

$$\mathbf{v}(\mathbf{x}, t) = \frac{d\psi_t}{dt}(\psi_t^{-1}(\mathbf{x})) \quad , \qquad (18)$$

satisfies a modified continuity equation. The proof is not given, since it consists in adapting an argument of Landau & Lifshitz [11, § 29] in the space-time M×**R**. The remaining arguments for obtaining Eq. (11) follow the lines of the classical derivation of d'Alembert's equation for the sound propagation and are less easy to check in the present context of a riemannian and time-dependent metric. However, d'Alembert's equation (in domains without massive bodies, i.e. Eq. (11) with nil right-hand side) means that pressure waves in the compressible ether propagate with a speed $c_e$ that is determined by the local ether compressibility K and obey the superposition principle : these features are purely local and should be conserved in the case of a time- and space-dependent metric. The fact that now $c_e = c$ and thus K is the constant $1/c^2$ is irrelevant when examining the validity of this equation. Since Eq. (11) (with nil time derivative) also is still applying in the steady situation, there does not seem to be other possible equation in the general case than the complete Eq. (11), with $c_e = c$ and thus $p_e = \rho_e c^2$ - and this is what we postulate. A further discussion is deferred until later, but it is important to note, as a consequence of our assumption (iv) of a space-contraction and time-dilatation in the ratio (12), that ether pressure (or density) waves are precisely waves of the space-time metric. Such waves are predicted by general relativity. In the present theory their appearance and their meaning are very clear.

5. SPACE-TIME METRIC AND GEODESIC ASSUMPTION

In its reinterpretation allowing for Lorentz contraction, the proposed gravitation theory consists in the field equation (11) with $p_e = \rho_e c^2$ for the (macro-) ether pressure, determining the slowing up of clocks and the contraction of measuring rods in the ratio (12), plus our modified Newton law (16) for the motion of a test particle in the ether. Note, however, that Eq. (16) makes sense only if the physical space metric is time-independent (in the frame E), for otherwise the time variation of the momentum cannot be defined as a vector. In general relativity, Newton's second law is replaced by Einstein's assumption that free test particles follow the geodesic lines of the "space-time metric". This latter measures simply the proper time $\tau$ along a trajectory (or line in space-time), i.e. the time measured with a clock bound with the considered test particle. The proper time, as compared with the time t measured with a fixed clock at rest in a given point $\mathbf{x}_0$ of ether, is affected by two effects : (a) since special relativity holds true locally, if the test particle has a velocity $\mathbf{u}$, with modulus u = |$\mathbf{u}$|, with respect to the momentarily coincident point $\mathbf{x}$ of ether (u is measured with clocks and rods of $\mathbf{x}$), $\tau$ is slowed up in the ratio

$$d\tau / dt_\mathbf{x} = \beta_\mathbf{u} = (1 - u^2/c^2)^{1/2} \quad , \quad (19)$$

with respect to the local time $t_\mathbf{x}$ ; and (b) the latter is affected in the ratio (Eq.(12)) :

$$dt_\mathbf{x} / dt = \rho_e(\mathbf{x}, t) / \rho_e(\mathbf{x}_0, t) = \rho_e / \rho_{e0} = \beta_g \quad , \quad (20)$$

(with the help of light signals from observers bound with ether, all the events may be referred to the time t of the point $\mathbf{x}_0$ of ether). The proper time of the test particle flows thus differently from the time t, in the ratio :

$$\frac{d\tau}{dt} = \frac{d\tau}{dt_\mathbf{x}} \frac{dt_\mathbf{x}}{dt} = \sqrt{(1 - u^2/c^2) \beta_g^2} = \sqrt{\beta_g^2 - (dl / dt_\mathbf{x})^2 (dt_\mathbf{x} / dt)^2 / c^2} \quad , \quad (21)$$

where $dl / dt_\mathbf{x} = u$ is measured with the physical space metric $\mathbf{g}$ : $u^2 = \mathbf{g}(\mathbf{u},\mathbf{u})$. Setting $x^0 = ct$, the space-time metric $\gamma$ is then defined by

$$ds^2 = c^2 d\tau^2 = \beta_g^2 (dx^0)^2 - dl^2 \quad , \quad (22)$$

in the sense that the modulus of the "4-velocity" of the test particle, $\mathbf{U} = (U^0, \mathbf{w})$ with $U^0 = dx^0 / ds$ and $\mathbf{w} = d\mathbf{x}/ds = \mathbf{u}\, dt_\mathbf{x}/ds$, is given by :

$$U^2 = \gamma(\mathbf{U}, \mathbf{U}) = \beta_g^2 (U^0)^2 - \mathbf{g}(\mathbf{w},\mathbf{w}) \quad . \quad (23)$$

Eq.(23) defines the square $U^2$ for any vector $\mathbf{U}$ in the tangent space at $(x^0, \mathbf{x})$ to the space-time $\mathbf{R} \times M$, although if $\mathbf{U}$ is not defined as the 4-velocity of an ordinary test particle or "time-like" line in space-time, $U^2$ may be negative or nil. The latter case corresponds to a light ray and the former to a "space-like" line in space-time.

It is proved in **Appendix 2** that *the trajectories of "free" test particles ($\mathbf{F}_0 = 0$) according to the modified Newton law (16) are geodesic lines of the space-time metric (22)*, with $dl$ as deduced from the euclidean metric by the space-contraction in the ratio (12) ; this result is valid for a constant gravitation field i.e. a constant ether density $\rho_e$. It is really a crucial result, for it at the same time establishes the necessary link between "classical" and reinterpreted ether theory and gives a firm basis to the gravitational space-contraction and time-dilatation in the ratio (12); here "classical" refers to the conception of uniform space and time. See Mazilu [9] for an assessment of the geodesic formulation in pure newtonian theory and in general relativity. In the unsteady situation where $\rho_e$ varies in time, there can be no Newton law any more and we *assume* by induction that the trajectories are geodesic lines of the metric (22). For an ordinary test particle, the geodesic assumption means that the interval of proper time along the followed line is *maximal* among the lines which join two given events $(x^0{}_1, \mathbf{x}_1)$ and $(x^0{}_2, \mathbf{x}_2)$. This is due to the nature of the indefinite pseudo-riemannian metric (22)-(23) : since we are speaking here of local geodesic lines, it may be verified for the flat Minkowski metric where the result is classical [4],[11].

An essential difference between the "classical" and reinterpreted versions of the present ether theory is that, in the former, the euclidean metric is assumed to be the physical one - otherwise the "classical" version, giving no standard of space-interval, would not be a physical theory ; in the same way, the "classical" version must neglect the influence of motion and gravitation on clocks.

## 6. CENTRAL STATIC SOLUTION REVISITED

As a fundamental test of the theory, let us determine the space-time metric in the static and spherically symmetrical situation already analysed in [1], but there with the euclidean interpretation of the field equations. Thus the mass-energy density $\rho$ as well as the ether pressure $p_e$ are time-independent and spherically symmetrical around the origin $\mathbf{x}=\mathbf{0}$. From assumption (iv), it follows that one may always select a system of curvilinear space coordinates $(x^i)$ such that $\partial p_e/\partial x^2 = \partial p_e/\partial x^3 = 0$, and with the natural basis $(\mathbf{e}_i)$ being orthogonal in the sense of both the euclidean metric $\mathbf{g}^0$ and the physical one $\mathbf{g}$ (see App. 2). Due to the spherical symmetry, we can evaluate the spatial variation of the metric with respect to "the" point of ether at $r = \infty$, where the two metrics must coincide : $g^0{}_{ij} = \delta_{ij} a_i{}^0$ and $g_{ij} = \delta_{ij} a_i{}^0 \lambda_i$ with $\lambda_1 = 1/\beta^2 = (\rho_e{}^\infty/\rho_e)^2$ and $\lambda_2 = \lambda_3 = 1$). Thus one has here :

$$\forall (dx^1, dx^2, dx^3), \quad \frac{dp_e}{dr} dr = \frac{dp_e}{dx^1} dx^1, \quad (24)$$

which implies that r depends only on $x^1$, hence $\mathbf{e}_1$ is a radial vector. Thus one may take here the spherical coordinates $x^1 = r$, $x^2 = \theta$, $x^3 = \phi$. With $d\Omega^2 = d\theta^2 + \sin^2\theta \, d\phi^2$, the spatial metric writes then :

$$dl^2 = (1/\beta^2) dr^2 + r^2 d\Omega^2, \quad (25)$$

and the space-time metric (Eq.(22)) is given by :

$$ds^2 = \beta^2 c^2 dt^2 - (1/\beta^2) dr^2 - r^2 d\Omega^2. \quad (26)$$

The grad operator (relative to $\mathbf{g}$) expresses for a scalar field f depending only on r :

$$\text{grad } f = g^{ij} \partial f/\partial x^j \mathbf{e}_i = \beta^2 \, df/dr \, \mathbf{e}_r, \quad \beta = \rho_e/\rho_e{}^\infty, \quad (27)$$

(where $g^{ij} = (\mathbf{g}^{-1})^{ij}$) and the div operator, applied to a such radial field $\mathbf{a} = a(r) \mathbf{e}_r$, is calculated as

$$\text{div } \mathbf{a} = (1/\sqrt{|g|})\, \partial(a^i \sqrt{|g|})/\partial x^i = (\beta/r^2)\, d/dr\, (r^2 a(r)/\beta)\,, \quad g = \det \mathbf{g} = r^4 \sin^2\theta\, /\beta^2. \quad (28)$$

Hence the field equation (11) (in this steady case, and with $p_e = \rho_e c^2$) writes

$$\Delta p_e \equiv \frac{\beta}{r^2}\frac{d}{dr}\left(r^2 \beta \frac{dp_e}{dr}\right) = A\, p_e\, \rho\,, \quad A = \frac{4\pi G}{c^2} \quad (29)$$

and with $\beta = \rho_e/\rho_e^\infty = p_e/p_e^\infty$ it takes the form :

$$\frac{d}{dr}\left(r^2 p_e \frac{dp_e}{dr}\right) = A(p_e^\infty)^2\, r^2\, \rho(r) \quad (30)$$

which by a first quadrature, involving the integration constant C, is equivalent to

$$\frac{d(p_e^2)}{dr} = 2\frac{C + A(p_e^\infty)^2 M(r)}{4\pi r^2}\,, \quad M(r) = 4\pi \int_0^r u^2 \rho(u)\, du. \quad (31)$$

This gives the gravitation field $\mathbf{g}$ : setting $\mathbf{g} = -g(r)\, \mathbf{e}_r$ obtains with Eqs. (10) and (27) :

$$g(r) = c^2\, (\text{grad } p_e)_r / p_e = c^2\, p_e \frac{dp_e}{dr} / (p_e^\infty)^2 \,, \quad (32)$$

thus Eq. (30) is just Poisson's equation for g(r). Inserting Eq. (32) into Eq. (31) gives

$$g(r) = \frac{G\, M(r)}{r^2} + \frac{C\, c^2}{4\pi r^2\, (p_e^\infty)^2} \,. \quad (33)$$

Assuming that $\rho(r)$ does not increase faster than $1/r$ as $r \to 0$, ensures that $M(r)/r^2$ remains bounded there. Then, just like in newtonian theory, the requirement that $g(r)$ remains bounded as $r \to 0$, implies that C=0 - and moreover *g(r) is then exactly the newtonian solution*. However this has not the same significance since here gravitation has the additional effect to alter the physical metric. Assuming that the whole mass-energy is finite : $M(r) \to M < \infty$, as $r \to \infty$, gives with Eq. (31) :

$$p_e^2 = -\frac{C}{2\pi r} - 2\frac{G(p_e^\infty)^2}{c^2}\mu(r) + D\,, \quad \mu(r) = \int_r^\infty \frac{M(u)}{u^2}\, du\,, \quad (34)$$

where the constant C was left. Eq. (34) informs us that, in the case where $\rho(r)$ would increase faster than $1/r$ as $r \to 0$, then either $\mu(r)$ would become infinite at r=0; in that case the constant C would have to be negative instead of nil, and further $\mu(r)$ would have to be at most in $1/r$ (otherwise the mathematical absurdity of a negative value $p_e^2$ would occur); or $\mu(r)$ would be bounded, in which case C again should be nil. In any case, $p_e^2$ is equivalent to D as $r \to \infty$, thus $p_e$ indeed becomes uniform at infinity and the non-euclidean character of the physical metric is measured by the value $\beta^2$. Returning to the "normal" situation where $M(r)/r^2$ is bounded (as well as $M(r)$) and C=0, we have :

$$\beta^2 = (p_e(r)/p_e^\infty)^2 = 1 - 2G\mu(r)/c^2 \,. \quad (35)$$

As r→ ∞, this admits the asymptotic expansion

$$\beta^2 = 1 - \frac{2GM}{c^2 r} + o\left(\frac{1}{r}\right). \quad (36)$$

The expansion (36) is exact, as soon as r ≥ R, if ρ(r) is assumed to cancel for r > R. This means that *the space-time metric* (Eq. (26)) :

$$ds^2 = \left(1 - \frac{2G\mu(r)}{c^2}\right) c^2 dt^2 - \frac{1}{1 - \frac{2G\mu(r)}{c^2}} dr^2 - r^2 d\Omega^2, \quad (37)$$

*is, for r ≥ R, exactly Schwarzschild's exterior metric (μ(r)=M/r) in the case where ρ(r)=0 for r > R*. Moreover this solution is asymptotically obtained at large r, in the case where ρ does not cancel outside a finite radius but the whole mass-energy is finite (this case is not considered in relativistic treatises). A striking difference with general relativity is that the whole solution (exterior and interior) is obtained at once and for a general mass distribution ρ(r), and furthermore *the interior solution does not coincide with Schwarzschild's interior solution* (the latter is obtained in general relativity when ρ is a constant for 0 ≤ r < R and nil for r > R). Indeed, the present theory yields then Eq.(37) with

$$\mu(r) = \frac{M}{R}\left(\frac{3}{2} - \frac{r^2}{2R^2}\right), \quad 0 \le r \le R, \quad (38)$$

whereas Schwarzschild's interior solution is :

$$ds^2 = \left(A - B\sqrt{1 - \frac{r^2}{R'^2}}\right)^2 c^2 dt^2 - \frac{1}{1 - \frac{r^2}{R'^2}} dr^2 - r^2 d\Omega^2, \quad (39)$$

$$1/R'^2 = 8\pi G \rho_0 / (3 c^2)$$

where the coefficients A and B are determined by assuming that the metric is continuous at r=R and that the kinetic pressure p cancels there [4]. Yet the most interesting difference with general relativity is perhaps the following one. In the case where ρ = 0 for r > R, the exact value $\beta^2 = 1 - 2GM/(c^2 r)$ for r ≥ R *precludes that the radius R of the spherical body is smaller than the "Schwarzschild radius" $r_S$* : one must have

$$R \ge 2GM/c^2 = r_S, \quad (40)$$

since otherwise a negative value $\beta^2 = (p_e / p_e^\infty)^2$ would occur at r=R. In general relativity, this inequality is in no way imposed. On the contrary, the relativistic analysis of gravitational collapse for very massive stars predicts the inevitable implosion of the star until a radius R < $r_S$ is reached, from which stage a bizarre situation is encountered, implying that R finally cancels, i.e. *the star becomes a singularity...*(e.g. [12]). *This cannot happen in the proposed theory*. To be complete, there is still the possibility that C < 0 ; this must be so if the mass-energy density increases so fast (as r→ 0) that μ(0) = +∞ and μ(r) = O(1/r) (see after Eq. (34)). This means that the newtonian behaviour (g(r) = GM/r²) is *not* recovered at large r, where the physical metric becomes euclidean : the field g (Eq. (33)) is indeed like 1/r² at large r, but with an active gravitational mass

$$M^a = g(r)\, r^2 / G = M + C\, c^2 /(4\pi G\, (p_e^\infty)^2) = M - m \qquad (41)$$

which is *smaller* than the inertial or passive gravitational mass M. Also, an unbounded repulsion occurs when r→ 0. Moreover the ether pressure $p_e$ tends towards infinity there (unless if μ(r) admits the asymptotic expansion : μ(r) = m/r + E + o(1) as r→ 0 ). In such cases (which seem wholly unphysical), the exterior metric would be Eq. (26) with

$$\beta^2 = \left(\frac{p_e}{p_e^\infty}\right)^2 = 1 - \frac{2G}{c^2 r}(M - m), \qquad (42)$$

i.e. a Schwarzschild metric with a smaller active mass $M^a < M$ would be obtained. In summary, if ρ(r)=0 for r > R, the exterior metric predicted by our theory is always a Schwarzschild metric, though perhaps with a smaller effective mass in very "pathological" cases, and in no case the Schwarzschild radius $2GM^a/c^2$ is greater than the radius R.

Thus the proposed theory is definitely not equivalent to general relativity. However it does produce Schwarzschild's exterior metric which supports all the essential tests of general relativity : the gravitational red shift of electromagnetic spectra, the delay of radar signals, the deviation of light rays and the advance in the perihelion of planetary orbits. By the way, this predicted advance, which agrees well with observations [12], is exactly six times greater than the advance which was predicted by our theory ([1], Eq. (60) where one assumes $c_e$=c) when the equations were first interpreted as relative to the euclidean metric. Surprisingly, the same underestimated prediction for this effect was obtained a long time ago by Poincaré [6] in the frame of his "electromagnetic" theory of gravitation, accounting for kinematic Lorentz contraction and assuming that gravitation propagates with the velocity of light. It is however clear that the basic principles and equations are very different in both theories.

6. CONCLUSION

In the view of Lorentz, Poincaré, Builder, Prokhovnik and others, the "relativistic" effects are in fact due to the absolute Lorentz contraction in the uniform motion with respect to the ether. Here the arguments for this interpretation of special relativity are presented afresh and briefly, but principally *gravitational space-contraction and time-dilatation* are postulated, based on the equivalence principle : in a theory based on ether, the absolute effects of uniform motion may be transposed into effects of a gravitational field i.e. a field of ether pressure. This is easily set into the most simple Eq. (12) *which implies the geodesic characterization of trajectories and a conservation of a potential plus kinetic energy* for a constant gravitation field. With the gravitational space-contraction, the basic equations of the proposed gravitation theory must be reinterpreted as relative to the "physical" metric, i.e. that which is affected by the contraction. Thus the proposed *unique* scalar field equation (11) has the mathematical peculiarity that the involved differential operator is relative to a metric deduced from the euclidean metric by a transformation which depends on the unknown function. However this in fact reduces to a different operator, this time relative to the euclidean metric.

For the spherical static problem, Schwarzschild's exterior space-time metric of general relativity is easily and exactly recovered, and with it all the experimental verifications of this latter theory. But differing from general relativity, the radius of the spherical body cannot be smaller than Schwarzschild's radius, and this is likely to prevent from any occurence of singularity - at least a "naked" one (the difficult question of gravitational collapse is a dynamical one and has not been discussed so far in the proposed theory). Moreover, the space-time metric, which is calculated at once and for a general spherical mass-energy distribution, is not Schwarzschild's interior metric inside a body with uniform mass-energy density. Another striking point is that *the gravity "acceleration" field of the spherical static problem is exactly the Newtonian one* (though it does not give the same motion since gravity alters the physical space-time metric also in the proposed theory).

The theory is not, and does not need to be formulated in a covariant way, since it assumes from the beginning a privileged frame (the *fluid* ether : the deformability of the ether is in terms of the *physical* metric, which differs from the euclidean one in the reinterpretation of the proposed theory allowing for Lorentz contraction). However it is fully consistent with the local Lorentz invariance and could clearly be transcribed in a covariant form, in the same way as a diagonal tensor may be rewritten in any base, thus filling the empty squares. Just in the spirit of Newton's theory, the motion of the ether with respect to a frame that is relevant for a given problem may be either postulated or empirically determined (as far as we today know, a large-scale expansion of the ether should give the correct description for all problems). Of course the author has not proved the existence of any ether, but perhaps he has contributed to make the assumption of an ether useful again.

ACKNOWLEDGEMENTS

I would like to gratefully thank Prof. P. Mazilu for helpful discussions. Moreover, Prof. Mazilu provided me with his relevant papers [9] before their publication. The interesting comments on the manuscript by Profs. P. Guélin and E. Soós are also kindly thanked.

**APPENDIX 1 : time derivative of a vector in a riemannian manifold**

The time derivative d**u**/dt of a vector **u**(t) attached to the point P(t) (**u**(t)$\in$ TM$_{P(t)}$ ) moving in the manifold M equipped with the metric **g** may be defined in the following way : for any vector **w** in the tangent space TM$_{P(t)}$ , let **w'**($\tau$) be the parallel transport of **w** on the trajectory $\tau \to P(\tau)$, relative to the metric **g**. Then d**u**/dt is the unique vector of TM$_{P(t)}$ which verifies :

$$\forall \mathbf{w} \in TM_{P(t)} \quad \frac{d\mathbf{u}}{dt} \cdot \mathbf{w} = \left(\frac{d}{d\tau}\right)_{\tau=t} (\mathbf{u}.\mathbf{w'}) \quad . \quad (A0)$$

where the scalar product is defined by **g**. This definition coincides with the usual one if (M, **g**) is euclidean (because then **w'**($\tau$) $\equiv$ **w**), and it is the only one which verifies

Leibniz's rule. Thus it is the right one (see below), but it is apparently not given in the literature.

**Theorem 1**. *Let M be a riemannian manifold and $t \to (P(t), \mathbf{u}(t))$ a differentiable mapping from an open interval $I \subset \mathbf{R}$ into the tangent bundle TM. Then Eq. (12) indeed defines, for any $t \in I$, a unique vector $\mathbf{y}=d\mathbf{u}/dt$ in the tangent space $TM_{P(t)}$, which is a time derivative in the sense of Leibniz's formula. Moreover this formula and the requirement that the derivative cancels for any parallel vector imply Eq. (12). The expression of $\mathbf{y}=d\mathbf{u}/dt$ in a given local system of coordinates is :*

$$y^m = u^m{}_{;t} = \frac{d u^m}{dt} + \Gamma^m_{ik} u^i v^k \quad , \quad (A1)$$

*where $\mathbf{v}=d\mathbf{P}/dt$ is the velocity vector of the trajectory $t \to P(t)$ and the $\Gamma$'s are the Christoffel symbols.*

This result justifies and precises the definition given by Eq. (A0), and proves that this time derivative is calculated (Eq.(A1)) just as the left-hand side of the equation of parallel transport which may thus be written $d\mathbf{u}/dt = 0$. It did not seem obvious, however, that the time derivative of a vector may be consistently defined as a vector in the general case and that it is given by Eq. (A1).

*Proof* . The right-hand side of Eq. (A0) defines a linear form $\mathbf{w} \to L(\mathbf{w})$ on $TM_{P(t)}$, because the parallel transport is a linear operation ; thus a unique vector $d\mathbf{u}/dt \in TM_{P(t)}$ is associated to this linear form by using the riemannian metric $\mathbf{g}$ ("raising up the indices"). Let $\mathbf{u}_1$ be another time-dependent vector on the same trajectory in M, i.e. the differentiable application $t \to (P(t), \mathbf{u}_1(t))$ from I to TM has the same projection P on M. Leibniz's formula :

$$\frac{d}{d\tau}\bigg|_{\tau=t} (\mathbf{u}.\mathbf{u}_1) = \frac{d\mathbf{u}}{dt} . \mathbf{u}_1(t) + \mathbf{u}(t).\frac{d\mathbf{u}_1}{dt} \quad , \quad (A2)$$

implies that if $\mathbf{u}_1 = \mathbf{w}'$ is parallel to itself on the trajectory P ($d\mathbf{u}_1/dt = 0$), then Eq.(A0) holds. Conversely, let us show that Leibniz's rule applies with the definition (A0). Selecting an orthonormal basis $(\mathbf{e}_{i0})$ in $TM_{P(t)}$ and transporting each vector $\mathbf{e}_i$ parallel to itself on the trajectory P : $\mathbf{e}_i = \mathbf{e}_{i0}'$, we obtain an orthonormal basis of $TM_{P(\tau)}$, since the scalar product of parallel vectors is conserved [8]. Then the definition (A0) gives $d\mathbf{e}_i/dt=0$ and Leibniz's rule obviously applies for any pair $(\mathbf{e}_i, \mathbf{e}_j)$. It is also easy to verify that the definition (A0) implies that $d/dt (a \mathbf{u}) = \mathbf{u} \, da/dt + a \, d\mathbf{u}/dt$ for any scalar function $a(t)$. Then the general rule (A2) follows as in the euclidean space. Let us calculate the contravariant coordinates $y^m$ of $\mathbf{y}=d\mathbf{u}/dt$ in a given coordinate system (we adopt the standard notations, see e.g. [8]). The time derivatives of the components of the parallel transport $\mathbf{w}'$ of $\mathbf{w}$ verify :

$$\dot{w}'^j = \frac{d w'^j}{d\tau}\bigg|_{\tau=t} = - \Gamma^j_{kl} w^k v^l \quad , \quad (A3)$$

where $v^l = dx^l/dt$ are the components of the velocity vector $\mathbf{v}$ ($x^l$ are the coordinates of P(t)). Developing Eq. (A0) with this, leads to

$$\forall (w^j), \ (g_{ij} \dot{u}^i + \dot{g}_{ij} u^i - g_{ik} u^i \Gamma^k_{jl} v^l - g_{ij} y^i) w^j = 0, \quad (A4)$$

thus at fixed j, the expression in parentheses cancels. Hence $y^m = g^{mj} g_{ij} y^i$ expresses :

$$y^m = \dot{u}^m - g^{mj} g_{ik} u^i \Gamma^k_{jl} v^l + g^{mj} \dot{g}_{ij} u^i. \quad (A5)$$

In Eq. (A5), the time variation of the metric is due only to the motion of P : $\dot{g}_{ij} = g_{ij,k} v^k$. Since the covariant derivative of the metric cancels :

$$g_{ij;k} = g_{ij,k} - \Gamma^l_{ik} g_{lj} - \Gamma^l_{jk} g_{il} = 0, \quad (A6)$$

it follows thus from Eq. (A5) that :

$$y^m - \dot{u}^m = g^{mj} (\Gamma^l_{ik} g_{lj} v^k u^i + \Gamma^l_{jk} g_{il} v^k u^i - g_{ik} \Gamma^k_{jl} u^i v^l), \quad (A7)$$

but since the sum of the two last terms in parentheses is nil, Eq. (A1) is proved.

**APPENDIX 2 : Geodesic formulation of the modified Newton law**

***Theorem 2.*** *If the gravitation field is constant in the frame E, then the solution trajectories of the modified Newton law with purely gravitational force :*

$$m(u) \mathbf{g} = \frac{d}{dt_\mathbf{x}} \left( m(u) \frac{d\mathbf{x}}{dt_\mathbf{x}} \right), \quad \frac{d}{dt_\mathbf{x}} \equiv \frac{\rho_{e0}}{\rho_e} \frac{d}{dt_{\mathbf{x}_0}} = \frac{p_{e0}}{p_e} \frac{d}{dt_{\mathbf{x}_0}}, \quad (A8)$$

*are geodesic lines of the space-time metric (22) involving the physical space metric $dl^2$ affected by the gravitational space-contraction in the ratio (12).*

*Proof.* Substituting the natural parameter s for $t_\mathbf{x}$ in (A8)$_1$ : $s = c\tau$ with $ds/dt_\mathbf{x} = c\beta_\mathbf{u} = c\,m(0)/m(u)$ (Eqs. (19) and (8)$_1$) obtains

$$m(u) \mathbf{g} = c^2 \frac{m(0)^2}{m(u)} \frac{d}{ds} \left( \frac{d\mathbf{x}}{ds} \right). \quad (A9)$$

The field $\mathbf{g}$ is given by Eq. (10) with $p_e = \rho_e c^2$ and the grad operator is relative to the physical space metric $dl^2$. Thus in a local coordinate system ($x^i$) where the bilinear form $\mathbf{g}$ associated with $dl^2$ has matrix ($g_{ij}$) with inverse matrix ($g^{ij}$) :

$$\mathbf{g} = -c^2 \frac{\text{grad } p}{p} = -c^2 \frac{g^{ij} p_{,j}}{p}, \quad (A10)$$

henceforth omitting the index e in p and ρ for simplicity. Using the expression (A1) of the "time" derivative d**v**/ds (with **v**=d**x**/ds), we get from (A9) and (A10) :

$$\frac{d^2 x^i}{ds^2} + \Gamma^i_{jk} \frac{dx^j}{ds} \frac{dx^k}{ds} + \frac{c^2}{c^2 - u^2} \frac{g^{ij} p_{,j}}{p} = 0 , \quad (A11)$$

where the (second-kind) Christoffel's Γ refer to the space metric **g**. We may select the coordinate system such that p=Const is equivalent to $x^1$=Const, and such that the euclidean metric **g**$^0$ is diagonal in the natural basis : $(g^0{}_{ij})$ = diag $(a^0_1, a^0_2, a^0_3)$. Assumption (iv) and Eq. (12) with p=ρ $c^2$ traduce in that **g** also is diagonal, and that :

$$(g_{ij}) = \text{diag} \left( \left(\frac{p^\infty}{p}\right)^2 a^0_1, a^0_2, a^0_3 \right) \equiv \text{diag}(a_i)_{1 \leq i \leq 3} , \quad (A12)$$

with $p^\infty = p(\mathbf{x}_0)$ ; $\mathbf{x}_0$ is a fixed point in M (i.e. $\mathbf{x}_0$ is at rest in the ether), outside the gravitation field (i.e. far enough). Introducing the time coordinate $x^0$=ct with $t=t_{\mathbf{x}_0}$, the space-time metric (22) writes in the local coordinates $(x^\alpha)$ (α=0,3) of the space-time **R**×M :

$$(\gamma_{\alpha\beta}) = \text{diag} \left( \left(\frac{p}{p^\infty}\right)^2, -\left(\frac{p^\infty}{p}\right)^2 a^0_1, -a^0_2, -a^0_3 \right) \equiv \text{diag}(b_\alpha)_{0 \leq \alpha \leq 3}$$

with $b_i = -a_i$ for $1 \leq i \leq 3$ ; henceforth, greek indices vary from 0 to 3 and latin ones from 1 to 3. From the general formula for the first-kind Christoffel symbols :

$$\Gamma_{ijk} = \frac{1}{2}(g_{ij,k} + g_{ik,j} - g_{jk,i}) \quad (A14)$$

it thus comes that the first-kind symbols Γ ' of the space-time metric verify :

$$\Gamma'_{ijk} = -\Gamma_{ijk} \quad i,j,k=1,2,3, \quad (A15)$$

$$\Gamma'_{\alpha\alpha\alpha} = \frac{1}{2} b_{\alpha,\alpha} , \Gamma'_{\alpha\beta\gamma} = 0 \quad \text{if } (\alpha \neq \beta \text{ and } \alpha \neq \gamma \text{ and } \beta \neq \gamma) \quad (A16)$$

$$\Gamma'_{\alpha\alpha\beta} = -\Gamma'_{\beta\alpha\alpha} = \frac{1}{2} \gamma_{\alpha\alpha,\beta} = \frac{1}{2} b_{\alpha,\beta} \quad \text{if } \alpha \neq \beta .$$

The assumption of a constant gravitational field means that the field p is independent on $t=t_{\mathbf{x}_0}$. So also are the coefficients $b_\alpha$ (Eq. (A13)) and we obtain from (A16) :

$$\Gamma'_{0\alpha\alpha} = \Gamma'_{\alpha 00} = \Gamma'_{\alpha 0\alpha} = 0, \quad \Gamma'_{\alpha 0\beta} = \Gamma'_{\alpha\beta 0} = \Gamma'_{0\alpha\beta} = 0 \text{ if } (\alpha \neq 0 \text{ and } \alpha \neq \beta \text{ and } \beta \neq 0) . \quad (A17)$$

Thus all $\Gamma'_{\alpha\beta\gamma}$ symbols with at least one index equal to zero cancel, except perhaps :

$$\Gamma'_{00i} = \Gamma'_{0i0} = -\Gamma'_{i00} = \frac{1}{2} b_{0,i} = p\, p_{,i} / (p^\infty)^2 \quad (A18)$$

which again is nil, unless if i=1. The second-kind Christoffel's are defined by

$$\Gamma^i_{jk} = g^{il}\Gamma_{ljk} = a_i^{-1}\Gamma_{ijk}, \quad \Gamma'^{\alpha}_{\beta\gamma} = b_\alpha^{-1}\Gamma'_{\alpha\beta\gamma}. \quad (A19)$$

Since $b_i = -a_i$ for $1 \le i \le 3$, we thus have from (A15):

$$\Gamma^i_{jk} = \Gamma'^i_{jk} \quad i,j,k = 1,2,3 \quad (A20)$$

from where the left-hand side of the geodesic equation writes for i=1,2,3:

$$G^i = \frac{d^2 x^i}{ds^2} + \Gamma'^i_{\alpha\beta}\frac{dx^\alpha}{ds}\frac{dx^\beta}{ds} = \frac{d^2 x^i}{ds^2} + \Gamma^i_{jk}\frac{dx^j}{ds}\frac{dx^k}{ds} + \Gamma'^i_{00}\left(\frac{dx^0}{ds}\right)^2 + 2\Gamma'^i_{0k}\frac{dx^0}{ds}\frac{dx^k}{ds}$$

(A21)

Thus we may insert the neo-newtonian Eq. (A11), in which we note that:

$$g^{ij}p_{,j} = a_i^{-1}p_{,i} = \begin{cases} 0 & \text{if } i \ne 1 \\ p_{,1}(p/p^\infty)^2 / a_1^0 & \text{if } i = 1 \end{cases} \quad (A22)$$

From (A22) and (A17), it follows that (A21) cancels for i=2,3, i.e. the geodesic equation is trivially satisfied for i=2,3. For i=1 (A21) reduces to

$$G^1 = -\frac{c^2}{c^2 - u^2}\frac{p\,p_{,1}}{a_1^0\,p^{\infty\,2}} + \Gamma'^1_{00}\left(\frac{dx^0}{ds}\right)^2. \quad (A23)$$

Now we have from (A18)-(A19) and (21) (and since $dx^0/ds = dt/d\tau$):

$$\Gamma'^1_{00} = \frac{p^3\,p_{,1}}{p^{\infty\,4}\,a_1^0}, \quad \left(\frac{dx^0}{ds}\right)^2 = \frac{p^{\infty\,2}}{p^2}\frac{c^2}{c^2 - u^2}. \quad (A24)$$

Hence the geodesic equation holds also for i=1. The last one, i.e. for i=0, writes in view of (A17) and (A19):

$$G^0 = \frac{d^2 x^0}{ds^2} + \Gamma'^0_{\alpha\beta}\frac{dx^\alpha}{ds}\frac{dx^\beta}{ds} = \frac{d}{ds}\frac{dx^0}{ds} + 2\Gamma'^0_{i0}\frac{dx^i}{ds}\frac{dx^0}{ds}. \quad (A25)$$

It comes from (A18) and (A19) that:

$$\Gamma'^0_{i0} = b_0^{-1}\Gamma'_{0\,i0} = p_{,i}/p \quad i=1,2,3 \quad (A26)$$

Using (A24)$_2$, we thus rewrite (A25), with $\gamma_u = 1/\sqrt{1 - u^2/c^2}$, as:

$$G^0 = \frac{d}{ds}\left(\frac{p^\infty}{p}\gamma_u\right) + 2\frac{p_{,1}}{p}\frac{dx^1}{ds}\frac{p^\infty}{p}\gamma_u = p^\infty\left[\left(\frac{1}{p}\frac{d\gamma_u}{ds}\right) - \frac{dp}{ds}\frac{\gamma_u}{p^2} + 2\frac{dp}{ds}\frac{\gamma_u}{p^2}\right]$$

$$= p^\infty\frac{\gamma_u}{p}\frac{d}{ds}(\text{Log}\,\gamma_u + \text{Log}\,p). \quad (A27)$$

The last equation will thus be obtained, and with it the proof of theorem 2, if we prove the following lemma which in itself is an essential result :

**Lemma.** *Any solution trajectory of the "free fall" motion (Eq. (A8)) admits the constant of motion* :

$$c^2 \, \mathrm{Log} \frac{1}{\sqrt{1 - u^2/c^2}} - U = \mathrm{Const} \,, \quad (A28)$$

where $U = - c^2 \, \mathrm{Log} \, p$ is *the potential [1], (38b) with* $p = \rho \, c^2$.

Thus with any constant gravitation field, we have the exact conservation of a potential plus kinetic energy - but the latter is *not* the kinetic energy $(8)_2$ of special relativity.

*Proof.* Extracting **g** = grad U from (A9) we get

$$\frac{dU}{ds} \equiv \mathrm{grad} \, U \cdot \frac{d\mathbf{x}}{ds} = (c^2 - u^2) \frac{d}{ds}\left(\frac{d\mathbf{x}}{ds}\right) \cdot \frac{d\mathbf{x}}{ds} = \frac{c^2 - u^2}{2} \frac{d}{ds}\left(\frac{d\mathbf{x}}{ds}\right)^2 \,, \quad (A29)$$

where the scalar products and square are defined with the metric **g**. Since $(ds/dt_\mathbf{x})^2 = c^2 (1 - u^2/c^2)$ (Eq. (19)) and since $u^2 = (d\mathbf{x}/dt_\mathbf{x})^2$, we have thus

$$\frac{dU}{ds} = \frac{c^2 - u^2}{2} \frac{d}{ds}\left(\frac{u^2}{c^2 - u^2}\right) = \frac{c^2 - u^2}{2} \frac{c^2}{(c^2 - u^2)^2} \frac{d(u^2)}{ds} = \frac{c^2}{2(c^2 - u^2)} \frac{d(u^2)}{ds} \,, \quad (A30)$$

which is also $(d/ds) \{c^2 \, \mathrm{Log} \, [ (1 - u^2/c^2)^{-1/2}] \}$. The proof is complete.

*Author's address*: Dr. M. ARMINJON, Fédération de Mécanique de Grenoble, Laboratoire "Sols, Solides, Structures", B. P. 53 X, F-38041 GRENOBLE cedex, France.